# Google Scholar Metrics 2014:
# a low cost bibliometric tool

**Alberto Martín-Martín[1], Juan Manuel Ayllón[1], Enrique Orduña-Malea[2], Emilio Delgado López-Cózar[1]**

[1] *EC3: Evaluación de la Ciencia y de la Comunicación Científica, Universidad de Granada (Spain)*
[2] *EC3: Evaluación de la Ciencia y de la Comunicación Científica, Universidad Politécnica de Valencia (Spain)*

## ABSTRACT

We analyse the main features of the third edition of Google Scholar Metrics (GSM), released in June 2014, focusing on its more important changes, strengths, and weaknesses. Additionally, we present some figures that outline the dimensions of this new edition, and we compare them to those of previous editions. Principal among these figures are the number of visualized publications, publication types, languages, and the maximum and minimum h5-index and h5-median values by language, subject area, and subcategory. This new edition is marked by continuity. There is nothing new other than the updating of the time frame (2009-2013) and the removal of some redundant subcategories (from 268 to 261) for English written publications. Google has just updated the data, which means that some of the errors discussed in previous studies still persist. To sum up, GSM is a minimalist information product with few features, closed (it cannot be customized by the user), and simple (navigating it only takes a few clicks). For these reasons, we consider it a "low cost" bibliometric tool, and propose a list of features it should incorporate in order to stop being labeled as such. Notwithstanding the above, this product presents a stability in its bibliometric indicators that supports its ability to measure and track the impact of scientific publications.

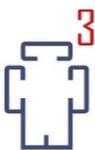

# 1. WHAT IS NEW IN GOOGLE SCHOLAR METRICS 2014?

No surprises. Almost with the punctuality of a fine Swiss watch, Google released, on June the 26th, 2014[1], its ranking of scientific publications: Google Scholar Metrics (GSM). Last year's version was published a bit later in the summer, on July the 24th, 2013. Google has stopped being different: it seems that from now on, these coveted lists of publications sorted by their scientific impact (that is, their h index) will be released every summer, whether it is on June, like this year, or July, like last year. This means that GSM: Google Scholar Metrics will join its competitors JCR (Journal Citation Reports) and SJR (Scimago Journal Rank) in updating the product on a yearly basis.

We can only welcome that the American company has decided to keep supporting GSM, a free product which is also very different from traditional journal rankings. This support might very well dispel the concerns about the continuity of its big brother, Google Scholar. Competition is healthy, and scientists can only be pleased about this variety of search and ranking tools, especially when they are offered free of charge.

Continuity and stability are the norm in this edition, since there is nothing new except for the removal of some subcategories (from 268 to 261) for English written publications. In short, Google has just updated the data, which means that some of the errors outlined in previous studies still persist [1-5]: the visualization of a limited number of publications (100 for those that are not published in English), the lack of categorization by subject areas and subcategories for non-English publications, and normalization problems (unification of journal titles, problems in the linking of documents, and problems in the search and retrieval of publication titles). As an example, it is inexcusable that there are duplicates to be found in a ranking of the top 100 publications (according to their h5-index) of a particular language. This is the case with the journal 日本機械学会論文集 C 編，which appears in the 42nd and 69th positions in the Japanese rankings, and in the Portuguese ranking, *Pesquisa Agropecuária Brasileira* appears in the 15th and 22nd positions, *Revista Brasileira de Engenharia Agrícola e Ambiental* in the 25th and 37th positions, and *Revista brasileira de atividade física e saúde* in the 91st and 93rd.

In our previous studies, we have described again and again the underlying philosophy embedded in all of Google's academic products. These products have been created in the image and likeness of Google's general search engine: fast, simple, easy to use, understand and, last but not least, accessible to everyone free of charge. GSM follows all these precepts, and it is, in the end, nothing more than:

- A hybrid between a bibliometric tool (indicators based on citation counts), and a bibliography (a list of highly cited documents, and of the documents that cite them).
- It offers a simple, straightforward journal classification scheme (although it also includes some conferences and repositories).

---

[1] http://googlescholar.blogspot.com.es/2014/06/2014-scholar-metrics-released.html

- It is based on two basic bibliometric indicators (the h index, and the median number of citations for the articles that contribute in the h index).
- It covers a single five-year time frame (the current one being 2009-2013).
- It uses rudimentary journal inclusion criteria, namely: publishing at least 100 articles during the last five-year period, and having received at least one citation.
- It provides lists of publications according to the language their documents are written in. For all of them, except for English publications (these are a total of 8: Chinese, Portuguese, German, Spanish, French, Japanese, Dutch, and Italian) it offers lists of only 100 titles: those with the higher h index. For English publications, however, it shows a total of 18 different publications, grouped in 8 subject areas and 261 subcategories. For each publication it shows the titles of the documents whose citations contribute to the h index, and for each one of these documents, in turn, the titles of the documents that cite them.
- It provides a search feature that, for any given set of keywords, will retrieve a list of 20 publications whose titles contain the selected keywords. In the cases where there are more than 20 publications that satisfy the query, only the first 20 results (those with a higher h index), will be displayed.
- It doesn't perform any kind of quality control in the indexing process nor in the information visualization process.

To sum up, GSM is a minimalist information product with few features, closed (it cannot be customized by the user), and simple (navigating it only takes a few clicks). If GSM wants to stop being labeled as a "low cost bibliometric tool", it should incorporate a wider range of features. At the very least, it should:

- Display the total number of publications indexed in GSM, as well as their countries and language of publication. Our estimations lead us to believe that this figure is probably higher than 40,000 [5]. In the case of Spain, there are over 1,000 publications indexed, which make up about 45% of the total number of academic publications in Spain [6, 7].
- Provide some other basic and descriptive bibliometric indicators, like the total number of documents published in the journals indexed in GSM, and the total number of citations received in the analysed time frame. These are the two essential parameters that make it possible to assess the reliability and accuracy of any bibliometric indicator. Other indicators could be added in order to elucidate other issues like self-citation rates, impact over time (immediacy index), or to normalize results (citation average).
- Provide the complete list of documents of any given publication that have received **n** citations and especially those that have received "0" citations. This would allow us to verify the accuracy of the information provided by this product. It is true, much to Google's credit, that this information could be extracted, though not easily, from Google Scholar.
- Provide a detailed list of the conferences and repositories included in the product. The statement Google makes about including some conferences in the Engineering & Computer Science area, and some document collections like the mega-repositories arXiv, RePec and SSRN, is much too vague.

- Define the criteria that has been followed for the creation of the classification scheme (areas and subcategories), and the rules and procedures followed when assigning publications to these areas and subcategories.
- Enable the selection of different time frames for the calculation of indicators and the visualization and sorting of publications. The significant disparities in publishing processes and citation habits between areas (publishing speed, pace of obsolescence) require the possibility to customize the time frame according to the particularities of any given subject area.
- Enable access to previous versions of Google Scholar Metrics (2007-2011, 2008-2012) to ensure that it is possible to assess the evolution of publications over time. Moreover, they could dare venture into the unknown and do something no one else has done before: a dynamic product, with indicators and rankings updated in real-time, just as Google Scholar does.
- Enable browsing publications by language, country and subcategory, and directly display all results for these selections.
- Remove visualization restrictions: currently 100 results for each language and 20 for each subcategory or keyword search.
- Enable the visualization of results by country of publication and by publisher.
- Enable sorting results according to various criteria (publication title, country, language, publishers), as well as according to other indicators (h index, h median, number of documents per publication, number of citations, self-citation rate…).
- Enable searching not only by publication title, but also by country and language of publication.
- Enable an option for exporting global results, as well as results by discipline, or those of a custom query.
- Enable an option for reporting errors detected by users, so they can be fixed (duplicate titles, erroneous titles, incorrect links, deficient calculations…).

## 2. GOOGLE SCHOLAR METRICS 2014 IN NUMBERS

Below we offer some data with the goal of shedding some light over the dimensions of this new edition of GSM and its differences, in quantitative terms, with previous versions of the product.

### 2.1. Publications visualized through GSM rankings

The total number of publications that can be visualized in the 2014 rankings is 7,100. Now, however, since 1,338 of them (24.7%) are classified in more than one subject area (Table 1), the number of unique publications is lower: 5,418.

***Table 1***
***Distribution of publications according to number of assigned subcategories***
***in Google Scholar Metrics 2014***

| Nº of subcategories | Publications | |
|---|---|---|
| | N | % |
| 1 | 4,100 | 75.7 |
| 2 | 1,131 | 20.9 |
| 3 | 125 | 2.3 |
| 4 | 51 | 0.9 |
| 5 | 6 | 0.1 |
| 6 | 5 | 0.1 |
| **Total** | **5,418** | **100.0** |

In 2014 there is a slight drop in the number of publications if we compare it with the previous edition (2008-2012), where 7,171 publications could be visualized in the rankings (5,462 unique publications).

## 2.2. Languages of visualized publications in GSM rankings

GSM rankings show an overwhelming skew towards the English language [8]: in this edition, the rankings show a total of 4,435 English-written publications (89%), and only 983 written in other languages, namely Chinese, Portuguese, German, Spanish, French, Italian, Japanese and Dutch (Figure 1). For all these languages, only 100 publications are shown in the language rankings, except in the case of Dutch publications, where only 83 items are shown.

***Figure 1***
***Distribution of publications according to languages***
***in the Google Scholar Metrics 2014 rankings***

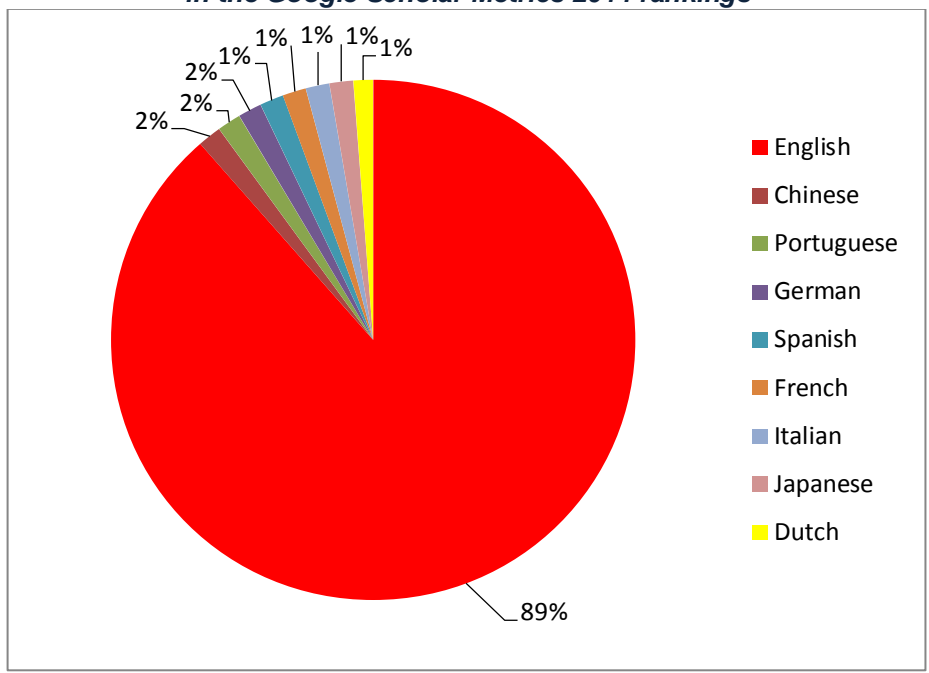

## 2.3. Types of publications visualized in GSM rankings

The predominant type of publication in the GSM rankings is the scientific journal: up to 5,068 of the total of 5,418 visualized publications (93.5%) belong to this type of publication (Figure 2).

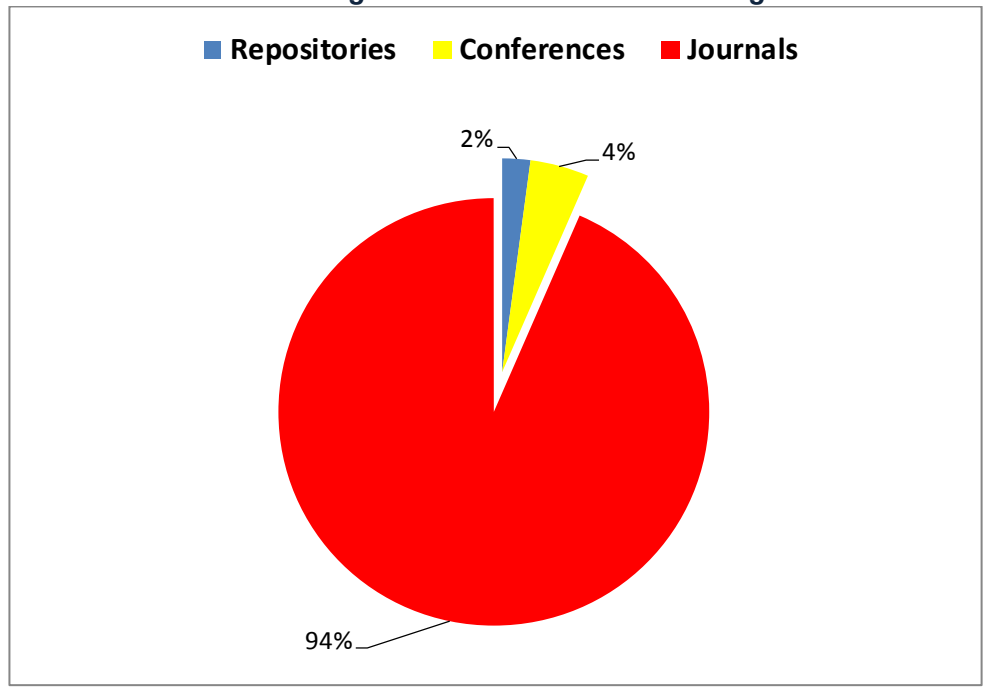

***Figure 2***
***Distribution of publications according to the type of source***
***in the Google Scholar Metrics 2014 rankings***

As is apparent from Table 2, if we compare the current version of the product with the last version (2008-2012), there haven't been any significant changes in the share each type of source holds.

***Table 2***
***Distribution of publications according to the type of source***
***in Google Scholar Metrics rankings 2008-2012 and 2009-2013***

| Sources | Publications | | | |
|---|---|---|---|---|
| | 2008-2012 | % | 2009-2013 | % |
| Repositories | 115 | 2.1 | 118 | 2.2 |
| Conferences | 241 | 4.4 | 232 | 4.3 |
| Journals | 5,107 | 93.5 | 5,068 | 93.5 |
| Total | 5,463 | 100 | 5,418 | 100 |

Conferences only make up 4.3% (232) of the visualized publications in the rankings. For the complete list of conferences, see Appendix 1. They are mostly concentrated in the Computer Science area, especially in subcategories like Computational Linguistics, Computer Hardware Design, Computer Networks & Wireless Communication, Computer Security & Cryptography, Computer Vision & Pattern Recognition, Computing Systems, Data Mining & Analysis, and Databases & Information Systems (Table 3).

The third type of publication indexed in GSM is the repository, whose inclusion has not been without certain controversy [9]. Repositories are digital storehouses with very wide subject coverage, created to save and disseminate very diverse academic materials. Because of the multidisciplinary nature of repositories, they shouldn't be compared to scientific journals and conferences, since each of the latter deals, generally, with much narrower fields of study (a subcategory or specialty) and because the documents they contain must endure a much stricter scientific review process in order to be selected for publication.

There are 118 visualized publications (2.2%) that belong to this type. Appendix 2 contains the complete list of visualized repositories. These publications are collections of documents that have been identified inside a repository, and they serve as the basis for the calculation of the bibliometric indicators. Most of them come from arXiv, RePec and SSRN, and are concentrated in the areas of Physics, Engineering and Economics (Table 3).

*Table 3*
*Distribution of publications according to type and subject area*
*Google Scholar Metrics rankings 2008-2012 and 2009-2013*

| Areas (English) | REPOSITORIES | | | | CONFERENCES | | | | JOURNALS | | | |
|---|---|---|---|---|---|---|---|---|---|---|---|---|
| | 2008-2012 | % | 2009-2013 | % | 2008-2012 | % | 2009-2013 | % | 2008-2012 | % | 2009-2013 | % |
| Business, Eco. & Manag. | 41 | 11.5 | 33 | 9.7 | 5 | 1.4 | 4 | 1.2 | 310 | 87.3 | 303 | 89.1 |
| Chemical & Material Sci. | 1 | 0.2 | 1 | 0.2 | 0 | 0.0 | 1 | 0.2 | 417 | 99.8 | 398 | 99.5 |
| Eng. & Computer Sci. | 23 | 1.9 | 25 | 2.1 | 244 | 20.4 | 229 | 19.4 | 929 | 77.7 | 926 | 78.5 |
| Health & Medical Sci. | 0 | 0.0 | 0 | 0.0 | 1 | 0.1 | 2 | 0.1 | 1,444 | 99.9 | 1,396 | 99.9 |
| Humanities, Lit. & Arts | 5 | 0.9 | 5 | 0.9 | 1 | 0.2 | 1 | 0.2 | 574 | 99.0 | 534 | 98.9 |
| Life Sciences & Earth Sci. | 6 | 0.7 | 7 | 0.9 | 3 | 0.4 | 4 | 0.5 | 823 | 98.9 | 789 | 98.6 |
| Physics & Mathematics | 78 | 14.8 | 78 | 15.6 | 21 | 4.0 | 19 | 3.9 | 429 | 81.2 | 403 | 80.6 |
| Social Sci. | 20 | 1.9 | 20 | 1.9 | 5 | 0.5 | 1 | 0.1 | 1,013 | 97.7 | 1,039 | 98.0 |

## 2.4. Classification scheme: areas and subcategories in GSM

As we pointed above, the only change introduced in the current version of GSM is the removal of a number of subcategories. There haven't been any changes in the eight broad subject areas (not in their designation nor in their size).

The areas that contain a larger number of subcategories are (Table 4): Health & Medical Sciences (69), Engineering & Computer Science (58) and Social Sciences (52). On the other hand, those with the lower number of subcategories are Business, Economics & Management (16) and Chemical & Material Sciences (19).

As can be seen in Table 4, the modifications introduced in this edition are the removal of a certain number of redundant subcategories across all areas, with the exception of Social Sciences, where we can actually find two more subcategories, which are: Bioethics and Sociology.

*Table 4*
*Number of subcategories by subject area*
*Google Scholar Metrics 2008-2012 and 2009-2013*

| Subject Areas | Nº of subcategories | |
|---|---|---|
| | 2008-2012 | 2009-2013 |
| Physics & Mathematics | 26 | 24 |
| Chemical & Material Sciences | 20 | 19 |
| Engineering & Computer Science | 59 | 58 |
| Health & Medical Sciences | 72 | 69 |
| Life Sciences & Earth Sciences | 41 | 39 |
| Humanities, Literature & Arts | 28 | 26 |
| Business, Economics & Management | 17 | 16 |
| Social Sciences | 50 | 52 |
| **TOTAL** | **287** | **269** |

It is important, however, to note that the number of unique subcategories is lower than the summation of subcategories in the eight aforementioned areas. This is so because some subcategories are included in more than one subject area. Particularly, this phenomenon is present in 44 subcategories in the 2008-2012 version of GSM, and in 42 subcategories in the current 2009-2013 version (Table 5). If these duplicates are not taken into account, the total number of unique subcategories for the 2008-2012 version is 268 subcategories, and 261 for the current version.

*Table 5*
*List of duplicate subcategories*
*Google Scholar Metrics 2008-2012 and 2009-2013*

| SUBCATEGORIES | 2008-2012 | 2009-2013 |
|---|---|---|
| Lipids | 3 | 0 |
| African Studies & History | 2 | 2 |
| Architecture | 2 | 2 |
| Asian Studies & History | 2 | 2 |
| Biochemistry | 2 | 2 |
| Bioethics | 2 | 2 |
| Bioinformatics & Computational Biology | 2 | 2 |
| Biomedical Technology | 2 | 2 |
| Biophysics | 2 | 2 |
| Biotechnology | 2 | 2 |
| Canadian Studies & History | 2 | 2 |
| Ceramic Engineering | 2 | 2 |
| Chinese Studies & History | 2 | 2 |
| Combustion & Propulsion | 2 | 2 |
| Development Economics | 2 | 2 |
| Economic History | 2 | 2 |
| Educational Administration | 2 | 2 |
| Educational Technology | 2 | 2 |
| Environmental & Geological Engineering | 2 | 2 |
| Epistemology & Scientific History | 2 | 2 |
| Feminism & Women's Studies | 2 | 2 |
| Food Science & Technology | 2 | 2 |
| Game Theory and Decision Science | 2 | 2 |
| History | 2 | 2 |
| Human Resources & Organizations | 2 | 2 |
| Library & Information Science | 2 | 2 |
| Materials Engineering | 2 | 2 |
| Medical Informatics | 2 | 2 |
| **Medicinal Chemistry** | 2 | 2 |
| **Microscopy** | 2 | 0 |
| **Middle Eastern & Islamic Studies** | 2 | 2 |
| **Molecular Biology** | 2 | 2 |
| **Molecular Modeling** | 2 | 2 |
| **Nanotechnology** | 2 | 2 |
| **Oil, Petroleum & Natural Gas** | 2 | 2 |
| **Paleontology** | 2 | 2 |
| **Plasma & Fusion** | 2 | 2 |
| **Public Health** | 2 | 2 |
| **Sex & Sexuality** | 2 | 2 |
| **Sustainable Development** | 2 | 2 |
| **Sustainable Energy** | 2 | 2 |
| **Technology Law** | 2 | 2 |
| **Virology** | 2 | 2 |
| **Wood Science & Technology** | 2 | 2 |
| **TOTAL** | **89** | **84** |

## 2.5. Bibliometric indicators of the publications in the GSM rankings

First of all, differences between scientific areas regarding bibliometric indicators become apparent once more. In Table 6 we present the maximum and minimum h5-index and h5-median values according to the subject area as shown in the GSM rankings for English written publications (Appendix 3 presents the same indicators by subcategories). It should be kept in mind that this is just a matter of the sizes and the cognitive and methodological natures of these scientific communities, which lead to very diverse publishing and citation habits.

*Table 6*
*Maximum and minimum h5-index and h5-median values according to the subject area in the GSM rankings for English written publications (2009-2013)*

| SUBJECT AREAS | Maximum H5-Index | Minimum H5-Index | Maximum H5-Median | Minimum H5-Median |
|---|---|---|---|---|
| **Life Sciences & Earth Sciences** | 355 | 108 | 495 | 146 |
| **Health & Medical Sciences** | 329 | 129 | 495 | 176 |
| **Chemical & Material Sciences** | 193 | 102 | 339 | 133 |
| **Physics & Mathematics** | 191 | 107 | 263 | 140 |
| **Engineering & Computer Science** | 174 | 93 | 253 | 130 |
| **Business, Economics & Management** | 168 | 63 | 241 | 92 |
| **Social Sciences** | 81 | 45 | 116 | 59 |
| **Humanities, Literature & Arts** | 38 | 25 | 72 | 30 |

Secondly, the significant differences among h index values for the publications of the nine languages covered in GSM (Tables 7a and 7b) should once again be stressed. English written publications have much better results than publications in other languages. It is no wonder that Chinese publications hold the second place, considering the high scientific output in this language.

It is surprising, however, that Portuguese publications seem to do better than Spanish publications, considering that the size of the Spanish-speaking scientific community is larger than the Portuguese community. The explanation may have something to do with Brazil's Open Access policies, exemplified by pioneer initiatives like Scielo.

*Table 7a*
*Maximun h5-index and h5-median values in the language rankings*
*Google Scholar Metrics: 2007-2011, 2008-2012 and 2009-2013 editions*

| LANGUAGE | H5-INDEX | | | H5-MEDIAN | | |
|---|---|---|---|---|---|---|
| | 2007-2011* | 2008-2012 | 2009-2013 | 2007-2011 | 2008-2012 | 2009-2013 |
| English | 121 | 349 | 355 | 174 | 504 | 495 |
| Chinese | 23 | 54 | 59 | 31 | 85 | 89 |
| Portuguese | 14 | 38 | 39 | 19 | 50 | 49 |
| Spanish | 11 | 28 | 27 | 16 | 38 | 37 |
| German | 10 | 22 | 27 | 15 | 28 | 37 |
| French | 7 | 18 | 15 | 10 | 19 | 18 |
| Korean | 5 | - | - | 7 | - | - |
| Japanese | 5 | 9 | 9 | 6 | 12 | 12 |
| Italian | 3 | 10 | 9 | 5 | 12 | 13 |
| Dutch | 2 | 10 | 6 | 5 | 12 | 8 |
| AVERAGE | 20 | 60 | 61 | 29 | 84 | 84 |

*April 2012 edition

*Table 7b*
*Minimum h5-index and h5-median values in the language rankings*
*Google Scholar Metrics: 2007-2011, 2008-2012 and 2009-2013 editions*

| LANGUAGE | H5-INDEX | | | H5-MEDIAN | | |
|---|---|---|---|---|---|---|
| | 2007-2011* | 2008-2012 | 2009-2013 | 2007-2011 | 2008-2012 | 2009-2013 |
| English | 88 | 105 | 107 | 112 | 140 | 142 |
| Chinese | 19 | 21 | 23 | 26 | 26 | 27 |
| Portuguese | 9 | 10 | 11 | 11 | 11 | 13 |
| Spanish | 8 | 10 | 10 | 10 | 12 | 12 |
| German | 6 | 7 | 7 | 8 | 8 | 13 |
| French | 6 | 6 | 7 | 6 | 7 | 8 |
| Korean | 4 | - | | 5 | - | |
| Japanese | 4 | 4 | 4 | 4 | 4 | 5 |
| Italian | 2 | 3 | 3 | 2 | 4 | 3 |
| Dutch | 1 | 1 | 1 | 1 | 1 | 1 |
| AVERAGE | 15 | 19 | 19 | 19 | 24 | 25 |

*April 2012 edition

It is also important to note the stability of these rankings. This can be considered from different points of view. On the one hand, there is the stability of the bibliometric indicators themselves. The differences between the first edition (2007-2011) and the two subsequent are due to the different publication dates (Tables 7a-b). The first edition was released in April 2012, while the second one was released in July 2013, and the current one in June 2014. Moreover, the reason why the results in the current edition (2009-2013) are so similar to those of the previous edition is probably related to the fact that it has been released one month ahead of last year's publication date (after eleven months, instead of twelve); however, a higher increase was expected, given the rapid growth Google Scholar shows [9,11].

In Table 8 we can see what the annual growth rates of the indicators have been for each of the broad subject areas, languages, and publication types. In the case of the language rankings, there are two different circumstances: h index values for English, Portuguese, Chinese and German publications (especially the last two) are increasing, while those of Spanish, French, Italian and Dutch are decreasing. The French and especially the Dutch cases draw our attention. Conversely, when we focus on the results by broad subject areas, we detect a more uniform behaviour: growth rates are mainly positive yet modest, except in the case of Physics & Mathematics. A similar situation can be found in the three publication types, although the higher growth in the group of conferences should be noted.

*Table 8*
*Variation in the maximum h5-index values by subject areas, languages and publication types. Google Scholar Metrics: 2008-2012 and 2009-2013 editions*

| SUBJECT AREAS | VARIATION % | |
|---|---|---|
| | H5-INDEX | H5-MEDIAN |
| Life Sciences & Earth Sciences | ▲ 1.72 | ▼ -1.79 |
| Health & Medical Sciences | ▲ 5.79 | ▼ -1.59 |
| Chemical & Material Sciences | ▲ 6.04 | ▲ 16.49 |
| Physics & Mathematics | ▼ -3.05 | ▲ 0.38 |
| Engineering & Computer Science | ▲ 6.75 | ▲ 5.86 |
| Business, Economics & Management | ▲ 4.35 | ▲ 8.56 |
| Social Sciences | ▲ 2.22 | ▲ 1.47 |
| Humanities, Literature & Arts | ▬ 0.00 | ▲ 10.77 |

| LANGUAGES | VARIATION % | |
|---|---|---|
| | H5-INDEX | H5-MEDIAN |
| English | ▲ 1.72 | ▼ -1.79 |
| Chinese | ▲ 9.26 | ▲ 4.71 |
| Portuguese | ▲ 2.63 | ▲ 4.00 |
| Spanish | ▼ -3.57 | ▲ 13.16 |
| German | ▲ 22.73 | ▲ 18.75 |
| French | ▼ -16.67 | ▼ -16.67 |
| Japanese | ▬ 0.00 | ▲ 7.69 |
| Italian | ▼ -10.00 | ▲ 13.33 |
| Dutch | ▼ -40.00 | ▲ 8.33 |

| PUBLICATION TYPES | VARIATION % | |
|---|---|---|
| | H5-INDEX | H5-MEDIAN |
| Journals | ▲ 1.72 | ▼ -1.79 |
| Conferences | ▲ 11.32 | ▼ -4.02 |
| Repositories | ▲ 4.35 | ▲ 8.56 |

On the other hand, the stability of GSM must be analysed from the perspective of the publications that enter or slide off the list, as well as from the perspective of the changes in the positions these publications hold in the rankings (Tables 9 and 10) when we compare the current version to the previous one (2008-2012).

Global rate of incoming/outgoing titles can be considered as very low (12.9%), although there are differences among languages. It is very low for English written publications (12.6%), and much higher for Chinese, Spanish, Italian and Japanese publications, where it goes above 25% (Table 9). If we apply this analysis to the publications grouped by subject areas –only for English written publications- the number of incoming/outgoing titles remains quite low (12.6% on average). Only in the areas of Humanities, Literature & Arts and Social Sciences do these values increase to 16% and 18% respectively (Table 10).

When we study the changes in publication ranks according to quartiles, we also find great stability: 71% of the publications stay in the same quartile as last year. However, there are differences among languages and subject areas. Thus, more than 50% of all French, Japanese and Italian publications find themselves in a different quartile as they were last year. Higher uniformity can be found in subject areas, where only in the areas of Humanities, Literature & Arts, and Social Sciences are these changes more abundant.

*Table 9*
***Changes in the list of publications visualized, and in their positions, in the Google Scholar Metrics language rankings (2009-2013)***

| LANGUAGES | ADDITIONS/ DELETIONS % | CHANGES IN PUBLICATION RANKS ACCORDING TO QUARTILES* | | |
|---|---|---|---|---|
| | | MOVES UP % | MOVES DOWN % | UNCHANGED % |
| English | 12.6 | 14.0 | 14.6 | 71.4 |
| Chinese | 25.0 | 18.7 | 20.0 | 61.3 |
| Portuguese | 15.0 | 23.2 | 20.7 | 56.1 |
| German | 19.0 | 19.8 | 19.8 | 60.5 |
| Spanish | 26.0 | 21.6 | 18.9 | 59.5 |
| French | 24.0 | 36.8 | 21.1 | 42.1 |
| Italian | 30.0 | 35.7 | 14.3 | 50.0 |
| Japanese | 30.0 | 21.7 | 30.4 | 47.8 |
| Dutch | 20.5 | 16.7 | 15.2 | 68.2 |
| AVERAGE | 12.9 | 14.3 | 14.7 | 71.0 |

* Common Publications in the two versions

*Table 10*
***Changes in the list of publications visualized, and in their positions, in the Google Scholar Metrics subject rankings (2009-2013)***

| SUBJECT AREAS (ENGLISH) | ADDITIONS/ DELETIONS % | CHANGES IN PUBLICATION RANKS ACCORDING TO QUARTILES* | | |
|---|---|---|---|---|
| | | MOVES UP % | MOVES DOWN % | UNCHANGED % |
| Business, Economics & Management | 10,6 | 11,2 | 14,7 | 74,1 |
| Chemical & Material Sciences | 13,2 | 12,7 | 14,2 | 73,0 |
| Engineering & Computer Science | 13,8 | 15,4 | 15,4 | 69,2 |
| Health & Medical Sciences | 10,9 | 12,4 | 13,6 | 74,0 |
| Humanities, Literature & Arts | 16,5 | 18,4 | 21,4 | 60,1 |
| Life Sciences & Earth Sciences | 12,3 | 15,2 | 13,6 | 71,2 |
| Physics & Mathematics | 10,2 | 10,2 | 11,9 | 77,9 |
| Social Sciences | 18,3 | 14,1 | 17,9 | 68,0 |
| AVERAGE | 12,6 | 14,0 | 14,6 | 71,4 |

* Common Publications in the two versions

Lastly, another test that can show us the degree of variability of the GSM rankings is the analysis of correlation between the ranks of the publications that are present in the two versions studied (2008-2012 and 2009-2013). If we calculate Pearson's correlation to the 4642 publications that are present in these two versions, the correlation is almost absolute: R=.99. A correlation analysis by languages and subcategories confirms the above. There is high correlation in almost the totality of cases. The lowest results can be found in the rankings of French, Japanese and Italian publications, and in the subcategories present in Humanities, Literature & Arts, and Social Sciences. Appendix 4 and 5 show the correlation results for each of the languages and subcategories present in GSM.

## 3. CONCLUSIONS

We have established that Google has opted for giving stability to its product instead of adding any more features or improvements, or even correcting some of its greatest shortfalls, which have been outlined in previous works (duplicate titles, errors in the links between documents and the source where they have been published, search deficiencies…). We should welcome the continuity of this initiative, which brings some fresh air to the scientific information market, and creates opportunities for the study of other kinds of impacts created by scientific output.

The stability of the indicators is a serious matter, since it reinforces the credibility of the data managed by Google Scholar in the construction of the GSM publication rankings. Despite all technical and methodological errors Google Scholar Metrics presents as a source for scientific evaluation, and despite the fact that we don't entirely know how everything works behind scenes, we can say that, from a stability point of view, these rankings can be considered as trustworthy and reliable.

This does not, however, prevent us from considering that there is clearly a lot of room for improvement for GSM as an information service. In analogy to the world of airlines, we cannot but consider GSM as a "low cost" bibliometric tool. To be completely fair, we should say "no cost", since users don't have to pay anything to access it (beyond the cost of the Internet connection), as opposed to what happens with real low cost airlines. From this point of view, it could be argued we shouldn't demand more of Google. Offering a tool that has required a great deal of effort for solving numerous technical problems (storing and processing great quantities of data, identifying different versions of the same document and merging them, calculating indicators, enabling fast and agile search and visualization features…) free of charge is definitely to be appreciated. However, we cannot overlook these technical and methodological shortcomings if we are to compare this product to other canonical bibliometric journal rankings.

Since it is based on the most comprehensive and less skewed scientific database in existance at the moment (Google Scholar), GSM is a valuable tool for learning the impact of thousands of publications that are not covered in any other bibliometric indexes. However, we cannot lose sight of its Achilles's heel, something we have been reporting insistently: it is vulnerable to tampering, and it is not transparent enough that the scientific community will be able to detect this kind of manipulation [10]. Publishers, in the pursuit of boosting their journals's impact, may not only manipulate editorial policies (which already happens more frequently than is generally thought), but they could also begin to directly control the impact of their journals by suggesting that authors cite studies published in the same journal. They may also, whether by themselves or in coordination with other parties, upload documents to the Web with citations to studies in their journal. For this reason, it is of the utmost importance that there should be two versions of the h index, one including self-citations, and one that does not take them into account. It is also necessary to find a way to clearly display the sources that provide citations.

## ACKNOWLEDGMENTS

This study has been funded through the project HAR2011-30383-C02-02 granted by the Dirección General de Investigación y Gestión del Plan Nacional de I+D+I, Ministerio de Economía y Competitividad, Spanish Government.

# APPENDIX 1
## Ranking of visualized conferences in Google Scholar Metrics (2009-2013)

| CONFERENCES | H5-INDEX | H5-MEDIAN |
|---|---|---|
| IEEE Conference on Computer Vision and Pattern Recognition, CVPR | 118 | 167 |
| International World Wide Web Conferences (WWW) | 83 | 131 |
| IEEE International Conference on Computer Vision, ICCV | 79 | 138 |
| Annual Joint Conference of the IEEE Computer and Communications Societies (INFOCOM) | 76 | 110 |
| ACM SIGKDD International Conference on Knowledge discovery and data mining | 69 | 113 |
| International Conference on Machine Learning (ICML) | 69 | 103 |
| ACM SIGCOMM Conference | 67 | 131 |
| EGU General Assembly Conference Abstracts | 66 | 100 |
| ACM Symposium on Information, Computer and Communications Security | 65 | 110 |
| International Conference on Very Large Databases | 64 | 92 |
| Meeting of the Association for Computational Linguistics (ACL) | 62 | 84 |
| ACM SIGMOD Conference | 61 | 90 |
| IEEE International Conference on Robotics and Automation | 61 | 87 |
| European Conference on Computer Vision | 59 | 96 |
| IEEE International Solid-State Circuits Conference | 59 | 79 |
| International Conference on Software Engineering | 57 | 94 |
| Society of Photo-Optical Instrumentation Engineers (SPIE) Conference Series | 56 | 70 |
| Symposium on Networked Systems: Design and Implementation (NSDI) | 54 | 84 |
| ACM International Conference on Web Search and Data Mining | 54 | 82 |
| International Symposium on Computer Architecture (ISCA) | 53 | 98 |
| IEEE Symposium on Security and Privacy | 53 | 85 |
| Conference on Empirical Methods in Natural Language Processing (EMNLP) | 53 | 78 |
| IEEE Power and Energy Society General Meeting | 53 | 77 |
| International Conference on Data Engineering Workshops | 52 | 68 |
| International Conference on Architectural Support for Programming Languages and Operating Systems (ASPLOS) | 51 | 87 |
| ACM Symposium on Theory of Computing | 51 | 77 |
| USENIX Security Symposium | 51 | 76 |
| ECS Meeting Abstracts | 51 | 75 |
| OFC/NFOEC Conference on Optical Fiber communication/National Fiber Optic Engineers Conference | 50 | 67 |
| IEEE International Symposium on Information Theory | 49 | 80 |
| ACM SIGIR Conference on Research and development in information retrieval | 49 | 71 |
| IEEE/IAS Industrial and Commercial Power Systems Technical Conference | 49 | 67 |
| International Conference on Weblogs and Social Media | 48 | 87 |
| Annual International Conference on Theory and Applications of Cryptographic Techniques (EUROCRYPT) | 48 | 79 |
| ACM International Conference on Information and Knowledge Management | 48 | 62 |
| Internet Measurement Conference | 47 | 78 |
| IEEE International Conference on Acoustics, Speech and Signal Processing (ICASSP) | 47 | 65 |
| IEEE Transactions on Circuits and Systems I: Regular Papers | 47 | 64 |
| Society for Economic Dynamics Meeting Papers | 46 | 80 |
| Annual International Conference on Mobile computing and networking | 46 | 72 |
| SIGPLAN Conference on Programming Language Design and Implementation (PLDI) | 46 | 71 |
| IEEE International Symposium on Parallel & Distributed Processing | 46 | 67 |
| International Joint Conference on Artificial Intelligence (IJCAI) | 46 | 63 |
| Conference on Advances in cryptology | 45 | 77 |
| IEEE GLOBECOM Workshops | 45 | 63 |
| IEEE/RSJ International Conference on Intelligent Robots and Systems | 45 | 62 |
| ACM SIGPLAN-SIGACT Symposium on Principles of Programming Languages (POPL) | 45 | 61 |
| IEEE International Electron Devices Meeting, IEDM | 45 | 60 |
| IEEE Symposium on Foundations of Computer Science (FOCS) | 45 | 58 |
| IEEE International Conference on Communications | 44 | 62 |
| ACM International Conference on Multimedia | 44 | 60 |
| IEEE Conference on Decision and Control | 43 | 63 |
| ACM SIAM Symposium on Discrete Algorithms | 43 | 59 |
| ACM European Conference on Computer Systems | 42 | 81 |
| IEEE International Conference on Cloud Computing (CLOUD) | 42 | 62 |
| IEEE International Symposium on High Performance Computer Architecture | 42 | 62 |
| IEEE Vehicular Technology Conference, VTC | 42 | 59 |
| Design, Automation and Test in Europe Conference and Exhibition (DATE) | 41 | 57 |
| IEEE/ACM International Symposium on Microarchitecture | 40 | 62 |
| ACM Symposium on User Interface Software and Technology | 40 | 62 |
| Design Automation Conference (DAC) | 40 | 54 |
| IEEE Energy Conversion Congress and Exposition (ECCE) | 40 | 53 |
| Network and Distributed System Security Symposium (NDSS) | 39 | 79 |
| National Fiber Optic Engineers Conference | 39 | 65 |
| International Conference on Spoken Language Processing (INTERSPEECH) | 39 | 52 |
| American Control Conference | 39 | 48 |
| IEEE Applied Power Electronics Conference and Exposition | 38 | 61 |
| International Conference on Computer Aided Verification (CAV) | 38 | 53 |
| European Semantic Web Symposium / Conference | 38 | 48 |
| Quantum Electronics and Laser Science Conference | 37 | 70 |
| Conference on File and Storage Technologies (FAST) | 37 | 62 |
| Conference on Computer Supported Cooperative Work (CSCW) | 37 | 60 |
| IEEE International Conference on Cloud Computing Technology and Science (CloudCom) | 37 | 58 |
| British Machine Vision Conference (BMVC) | 37 | 56 |
| AAAI Conference on Artificial Intelligence | 37 | 50 |
| Conference on Lasers and Electro-Optics | 36 | 63 |
| ACM Conference on Recommender Systems | 36 | 62 |
| International Conference on Distributed Computing Systems, ICDCS | 36 | 49 |
| IEEE International Conference on Data Mining (ICDM) | 36 | 47 |
| International Conference on Extending Database Technology (EDBT) | 36 | 47 |
| ACM SIGPLAN Symposium on Principles & Practice of Parallel Programming (PPOPP) | 35 | 69 |
| USENIX Annual Technical Conference | 35 | 54 |
| SIAM International Conference on Data Mining | 35 | 49 |
| IEEE International Conference on Image Processing (ICIP) | 35 | 47 |
| International Conference on The Theory and Application of Cryptology and Information Security (ASIACRYPT) | 34 | 58 |
| IEEE International Symposium on Cluster Computing and the Grid | 34 | 50 |
| Conference on Object-Oriented Programming Systems, Languages, and Applications (OOPSLA) | 34 | 50 |
| IEEE International Conference on Computer Vision Workshops (ICCV Workshops) | 33 | 59 |
| Allerton Conference on Communication, Control, and Computing | 33 | 49 |
| Workshop on Cryptographic Hardware and Embedded Systems (CHES) | 33 | 45 |
| International Quantum Electronics Conference | 32 | 57 |
| ACM Conference on Electronic Commerce | 32 | 47 |
| IEEE Congress on Evolutionary Computation | 32 | 39 |
| International Conference on Computational Linguistics (COLING) | 31 | 48 |
| International Symposium on Software Testing and Analysis | 31 | 47 |
| International Conference on Tools and Algorithms for the Construction and Analysis of Systems (TACAS) | 31 | 44 |
| International Symposium/Conference on Music Information Retrieval | 31 | 43 |
| International Conference on Document Analysis and Recognition | 31 | 41 |
| Conference on Genetic and Evolutionary Computation | 31 | 39 |
| International Conference on Practice and Theory in Public Key Cryptography | 30 | 53 |
| IEEE Real-Time Systems Symposium (RTSS) | 30 | 53 |
| Conference of the European Chapter of the Association for Computational Linguistics (EACL) | 30 | 45 |
| IEEE/ACM International Conference on Automated Software Engineering (ASE) | 30 | 44 |
| IEEE International Conference on Software Testing, Verification and Validation Workshops (ICSTW) | 30 | 38 |
| Computer Security Applications Conference | 29 | 50 |
| Symposium On Usable Privacy and Security | 29 | 47 |
| ACM/IEEE International Conference on Human Robot Interaction | 29 | 45 |
| ACM Symposium on Parallelism in Algorithms and Architectures (SPAA) | 29 | 44 |
| IEEE/ACM International Conference on Computer-Aided Design (ICCAD) | 29 | 39 |
| Symposium on Theoretical Aspects of Computer Science (STACS) | 29 | 38 |
| IFAC World Congress | 29 | 35 |
| International Conference on Pattern Recognition | 29 | 35 |
| European Conference on Research in Computer Security | 28 | 49 |
| European Conference on Object-oriented Programming (ECOOP) | 28 | 45 |
| International Conference on Functional Programming (ICFP) | 28 | 42 |
| International Conference on Financial Cryptography and Data Security | 28 | 40 |
| Asia and South Pacific Design Automation Conference (ASP-DAC) | 28 | 36 |
| IEEE Information Theory Workshop | 28 | 36 |
| IEEE Computer Security Foundations Symposium | 27 | 40 |
| European Conference on Machine learning and knowledge discovery in databases | 27 | 40 |
| ACM International Conference on Interactive Tabletops and Surfaces (ITS) | 27 | 39 |
| IEEE Conference of Industrial Electronics | 27 | 39 |
| International Conference on Biometrics | 27 | 34 |
| IEEE International Symposium on Circuits and Systems | 27 | 33 |
| IEEE/MTT-S International Microwave Symposium | 27 | 32 |
| Symposium on Interactive 3D Graphics (SI3D) | 26 | 38 |
| International Conference on Intelligent User Interfaces (IUI) | 26 | 38 |
| Conference on Information Sciences and Systems | 26 | 38 |
| Symposium on Field Programmable Gate Arrays (FPGA) | 26 | 37 |
| IEEE Vehicle Power and Propulsion Conference | 26 | 37 |
| ACM SIGGRAPH/Eurographics Symposium on Computer Animation | 26 | 36 |
| IEEE Symposium on Logic in Computer Science | 26 | 36 |
| Electronic Components and Technology Conference, ECTC | 26 | 35 |
| IEEE PES Power Systems Conference and Exposition | 25 | 38 |
| Asian Conference on Computer Vision | 25 | 36 |
| European Conference on Algorithms | 25 | 35 |
| Symposium on Computational Geometry | 25 | 35 |
| IEEE Conference on Computational Intelligence and Games | 25 | 34 |
| European Quantum Electronics Conference | 24 | 50 |
| IEEE International Reliability Physics Symposium (IRPS) | 24 | 41 |
| Information Theory and Applications Workshop | 24 | 39 |
| Workshop of Cross-Language Evaluation Forum | 24 | 39 |
| ACM/IEEE International Symposium on Low Power Electronics and Design | 24 | 38 |
| International Conference on Affective Computing and Intelligent Interaction and Workshops | 24 | 36 |
| ACM/IEEE International Symposium on Networks-on-Chip | 24 | 35 |
| IEEE Intelligent Vehicles Symposium | 24 | 33 |
| International Conference on Tangible, embedded, and embodied interaction | 24 | 31 |
| European Conference on Power Electronics and Applications | 24 | 31 |
| IFIP Conference on Human-Computer Interaction (INTERACT) | 24 | 28 |
| IEEE Custom Integrated Circuits Conference, CICC | 23 | 33 |
| IEEE Radio Frequency Integrated Circuits Symposium | 23 | 33 |
| IEEE International Conference on Advanced Video and Signal-Based Surveillance (AVSS) | 23 | 33 |
| IEEE-RAS International Conference on Humanoid Robots (Humanoids) | 23 | 33 |
| Pacific Symposium on Biocomputing | 23 | 30 |
| IEEE International Conference on Multimedia and Expo | 23 | 28 |
| Pacific-Asia Conference on Advances in Knowledge Discovery and Data Mining | 22 | 34 |
| International Conference on Research in Computational Molecular Biology | 22 | 32 |
| ACM Multimedia Systems Conference (MMSys) | 22 | 31 |
| International ACM/IEEE Joint Conference on Digital Libraries | 21 | 33 |
| IEEE Radar Conference | 21 | 31 |
| IEEE International Conference on RFID | 21 | 27 |
| IEEE International Conference on Fuzzy Systems (FUZZ) | 21 | 26 |
| International Workshop on Semantic Evaluation | 20 | 49 |
| Eurographics - Short Papers | 20 | 38 |
| IEEE Symposium on VLSI Circuits | 20 | 29 |
| Picture Coding Symposium (PCS) | 20 | 29 |
| Annual Meeting of the Special Interest Group on Discourse and Dialogue (SIGDIAL) | 20 | 28 |
| IEEE International Geoscience and Remote Sensing Symposium | 20 | 28 |
| Workshop on Statistical Machine Translation | 20 | 27 |
| IEEE Pacific Visualization Symposium | 20 | 27 |
| IEEE Workshop on Automatic Speech Recognition & Understanding | 20 | 27 |
| European Conference on Antennas and Propagation | 20 | 26 |
| IEEE Conference on Emerging Technologies & Factory Automation (ETFA) | 20 | 24 |
| Conference on Image and Video Retrieval | 19 | 33 |
| International Workshop on Quality of Multimedia Experience | 19 | 31 |
| IUFRO Workshop on Connection Between Forest Resources and Wood Quality: Modeling Approaches and Simulation Software | 19 | 29 |
| European Conference on Evolutionary Computation, Machine Learning and Data Mining in Bioinformatics | 19 | 26 |
| IEEE Aerospace Conference | 19 | 25 |
| AMIA Symposium | 19 | 25 |
| International Electron Devices Meeting | 19 | 24 |
| International Conference on Indoor Positioning and Indoor Navigation (IPIN) | 19 | 24 |
| IEEE Workshop on Signal Processing Advances in Wireless Communications | 18 | 27 |
| International Symposium on Software Reliability Engineering | 18 | 25 |
| IEEE Workshop on Applications of Signal Processing to Audio and Acoustics | 18 | 23 |
| International Workshop on Internet and Network Economics | 17 | 30 |
| IEEE International Conference on Industrial Informatics | 17 | 24 |
| International Offshore and Polar Engineering Conference | 17 | 23 |
| World Congress on Nature & Biologically Inspired Computing, NaBIC | 17 | 22 |
| International Conference on Machine Learning and Cybernetics | 17 | 22 |
| International Conference on Ocean, Offshore and Arctic Engineering | 17 | 20 |
| IEEE/AIAA Digital Avionics Systems Conference (DASC) | 16 | 26 |
| Annual Conference on Computer Graphics (SIGGRAPH) | 16 | 26 |
| ACM/IEEE International Conference on Distributed Smart Cameras | 16 | 23 |
| International Conference on Computational Linguistics and Intelligent Text Processing | 16 | 22 |
| International Conference on Parallel problem solving from nature | 16 | 21 |
| Linguistic Annotation Workshop | 16 | 20 |
| IEEE Symposium on Computational Intelligence and Data Mining | 15 | 30 |
| Data Compression Conference, DCC | 15 | 26 |
| IEEE International Test Conference | 15 | 22 |
| International Joint Conference on Natural Language Processing | 15 | 21 |
| ENTER eTourism Conference | 15 | 20 |
| Workshop on Positioning, Navigation and Communication | 15 | 20 |
| IEEE International Conference on Semantic Computing | 15 | 18 |
| European Conference on Genetic Programming | 15 | 18 |
| IEEE Spoken Language Technology Workshop (SLT) | 15 | 17 |
| International Conference on Natural Computation | 15 | 17 |
| Web3D / VRML Symposium | 14 | 20 |
| IEEE/ION Position, Location and Navigation Symposium | 14 | 19 |
| European Conference on Synthetic Aperture Radar | 14 | 19 |
| International Conference on Fuzzy Systems and Knowledge Discovery | 14 | 18 |
| ACM Symposium on Virtual Reality Software and Technology | 14 | 17 |
| ACM International Health Informatics Symposium | 14 | 17 |
| International Conference on Computer Safety, Reliability, and Security | 14 | 17 |
| IEEE International Conference on Ultra-Wideband | 14 | 17 |
| IEEE International Conference on Plasma Science (ICOPS) | 13 | 31 |
| International ITG Workshop on Smart Antennas (WSA) | 13 | 21 |
| European Radar Conference | 13 | 16 |
| IEEE Electric Ship Technologies Symposium | 12 | 20 |
| International Symposium on Turbo Codes and Iterative Information Processing | 12 | 17 |
| GI-Jahrestagung | 11 | 14 |
| Multikonferenz Wirtschaftsinformatik (MKWI) | 10 | 15 |
| International Radar Symposium | 10 | 12 |
| International Conference on Quantum Information | 9 | 21 |
| Biennial Symposium on Communications (QBSC) | 9 | 17 |
| International Conference on Computer Graphics, Imaging and Visualization, CGIV | 9 | 14 |
| IEEE International Symposium on Phased Array Systems and Technology | 9 | 12 |
| IEEE/NPSS Symposium on Fusion Engineering SOFE | 9 | 11 |
| Australian Communications Theory Workshop (AusCTW) | 8 | 10 |
| Canadian Workshop on Information Theory | 8 | 10 |
| International Symposium on Information Theory and Its Applications | 8 | 10 |
| Asian-Pacific Conference on Synthetic Aperture Radar | 8 | 9 |
| International Conference on Ground Penetrating Radar | 7 | 11 |
| ESA Workshop on Satellite Navigation Technologies and European Workshop on GNSS Signals and Signal Processing (NAVITEC) | 7 | 9 |
| Conférence en Recherche d'Infomations et Applications (CORIA) | 7 | 9 |
| IEEE CIE International Conference on Radar | 6 | 7 |

# APPENDIX 2

## Ranking of repositories visualized in Google Scholar Metrics (2009-2013)

| REPOSITORIES | H5-INDEX | H5-MEDIAN |
|---|---|---|
| NBER Working Papers | 168 | 241 |
| arXiv Cosmology and Extragalactic Astrophysics (astro-ph.CO) | 162 | 222 |
| arXiv High Energy Physics - Phenomenology (hep-ph) | 145 | 194 |
| arXiv Mesoscale and Nanoscale Physics (cond-mat.mes-hall) | 138 | 198 |
| arXiv Materials Science (cond-mat.mtrl-sci) | 136 | 227 |
| arXiv Quantum Physics (quant-ph) | 128 | 204 |
| arXiv High Energy Physics - Theory (hep-th) | 124 | 174 |
| arXiv High Energy Physics - Experiment (hep-ex) | 118 | 178 |
| arXiv Superconductivity (cond-mat.supr-con) | 109 | 162 |
| arXiv High Energy Astrophysical Phenomena (astro-ph.HE) | 107 | 140 |
| CEPR Discussion Papers | 105 | 148 |
| arXiv Galaxy Astrophysics (astro-ph.GA) | 101 | 138 |
| arXiv Solar and Stellar Astrophysics (astro-ph.SR) | 101 | 127 |
| arXiv Earth and Planetary Astrophysics (astro-ph.EP) | 95 | 128 |
| arXiv Information Theory (cs.IT) | 92 | 137 |
| arXiv Quantum Gases (cond-mat.quant-gas) | 91 | 139 |
| arXiv Strongly Correlated Electrons (cond-mat.str-el) | 90 | 129 |
| arXiv Optics (physics.optics) | 90 | 125 |
| IZA Discussion Papers | 81 | 118 |
| arXiv Nuclear Theory (nucl-th) | 78 | 101 |
| arXiv Physics and Society (physics.soc-ph) | 76 | 109 |
| arXiv General Relativity and Quantum Cosmology (gr-qc) | 76 | 98 |
| arXiv Instrumentation and Methods for Astrophysics (astro-ph.IM) | 71 | 115 |
| arXiv High Energy Physics - Lattice (hep-lat) | 69 | 94 |
| arXiv Statistical Mechanics (cond-mat.stat-mech) | 69 | 94 |
| arXiv Nuclear Experiment (nucl-ex) | 68 | 100 |
| Policy Research Working Paper Series | 62 | 87 |
| MPRA Paper | 62 | 82 |
| arXiv Statistics Theory (math.ST) | 61 | 102 |
| arXiv Optimization and Control (math.OC) | 61 | 99 |
| arXiv Soft Condensed Matter (cond-mat.soft) | 59 | 81 |
| arXiv Atomic Physics (physics.atom-ph) | 58 | 90 |
| European Central Bank Working Paper Series | 57 | 98 |
| CESifo Working Paper Series | 57 | 87 |
| arXiv Other Condensed Matter (cond-mat.other) | 52 | 77 |
| arXiv Learning (cs.LG) | 50 | 78 |
| arXiv Probability (math.PR) | 50 | 61 |
| arXiv Instrumentation and Detectors (physics.ins-det) | 49 | 84 |
| CEP Discussion Papers | 48 | 85 |
| arXiv Mathematical Physics (math-ph) | 47 | 65 |
| Bank for International Settlements Working Papers | 46 | 97 |
| arXiv Networking and Internet Architecture (cs.NI) | 46 | 72 |
| arXiv Analysis of PDEs (math.AP) | 45 | 54 |
| arXiv Disordered Systems and Neural Networks (cond-mat.dis-nn) | 44 | 66 |
| arXiv Data Structures and Algorithms (cs.DS) | 44 | 56 |
| arXiv Methodology (stat.ME) | 43 | 65 |
| arXiv Fluid Dynamics (physics.flu-dyn) | 43 | 54 |
| arXiv Databases (cs.DB) | 42 | 57 |
| Bank of Italy Working Papers | 41 | 63 |
| arXiv Machine Learning (stat.ML) | 40 | 68 |
| eSocialSciences Working Papers | 40 | 61 |
| arXiv Plasma Physics (physics.plasm-ph) | 40 | 57 |
| arXiv Populations and Evolution (q-bio.PE) | 40 | 56 |
| arXiv Biological Physics (physics.bio-ph) | 40 | 50 |
| arXiv Distributed, Parallel, and Cluster Computing (cs.DC) | 39 | 71 |
| arXiv Numerical Analysis (math.NA) | 39 | 49 |
| World Bank Policy Research Working Paper Series | 38 | 63 |
| arXiv Computer Vision and Pattern Recognition (cs.CV) | 38 | 50 |
| arXiv Algebraic Geometry (math.AG) | 38 | 48 |
| Department of Economics and Business, Universitat Pompeu Fabra Economics Working Papers | 37 | 72 |
| arXiv Combinatorics (math.CO) | 37 | 46 |
| HAL Working Papers | 36 | 69 |
| arXiv Social and Information Networks (cs.SI) | 36 | 60 |
| arXiv Applications (stat.AP) | 36 | 52 |
| arXiv Digital Libraries (cs.DL) | 35 | 65 |
| arXiv Number Theory (math.NT) | 34 | 48 |
| Harvard Business School Working Papers | 33 | 61 |
| Harvard University, John F. Kennedy School of Government Working Paper Series | 33 | 59 |
| arXiv Functional Analysis (math.FA) | 33 | 52 |
| arXiv Chaotic Dynamics (nlin.CD) | 33 | 46 |
| arXiv Computer Science and Game Theory (cs.GT) | 33 | 44 |
| arXiv Differential Geometry (math.DG) | 33 | 43 |
| Federal Reserve Bank of San Francisco Working Paper Series | 32 | 73 |
| OECD Economics Department Working Papers | 32 | 56 |
| arXiv Quantitative Methods (q-bio.QM) | 32 | 49 |
| Deep Sea Research Part I: Oceanographic Research Papers | 32 | 44 |
| arXiv Computational Complexity (cs.CC) | 32 | 39 |
| arXiv Cryptography and Security (cs.CR) | 31 | 50 |
| CReAM Discussion Paper Series | 31 | 50 |
| arXiv Pattern Formation and Solitons (nlin.PS) | 31 | 41 |
| IFPRI discussion papers | 30 | 50 |
| arXiv Representation Theory (math.RT) | 30 | 37 |
| arXiv Exactly Solvable and Integrable Systems (nlin.SI) | 29 | 45 |
| arXiv Genomics (q-bio.GN) | 29 | 44 |
| arXiv Logic in Computer Science (cs.LO) | 29 | 43 |
| arXiv Molecular Networks (q-bio.MN) | 29 | 41 |
| arXiv Classical Analysis and ODEs (math.CA) | 29 | 39 |
| arXiv Dynamical Systems (math.DS) | 29 | 35 |
| Hong Kong Institute for Monetary Research Working Papers | 28 | 42 |
| Resources For the Future Discussion Papers | 28 | 40 |
| arXiv Computation and Language (cs.CL) | 27 | 57 |
| arXiv Biomolecules (q-bio.BM) | 27 | 44 |
| arXiv Programming Languages (cs.PL) | 27 | 41 |
| World Institute for Development Economic Research (UNU-WIDER) Working Paper Series | 26 | 38 |
| arXiv Discrete Mathematics (cs.DM) | 26 | 30 |
| arXiv Group Theory (math.GR) | 25 | 31 |
| arXiv Geometric Topology (math.GT) | 24 | 34 |
| arXiv Quantum Algebra (math.QA) | 24 | 32 |
| arXiv Operator Algebras (math.OA) | 24 | 30 |
| Human Development Research Papers | 23 | 41 |
| arXiv Complex Variables (math.CV) | 23 | 32 |
| arXiv Symplectic Geometry (math.SG) | 23 | 28 |
| arXiv Rings and Algebras (math.RA) | 23 | 27 |
| arXiv Commutative Algebra (math.AC) | 22 | 26 |
| arXiv Atomic and Molecular Clusters (physics.atm-clus) | 21 | 33 |
| arXiv Algebraic Topology (math.AT) | 21 | 30 |
| arXiv Metric Geometry (math.MG) | 20 | 30 |
| arXiv Computational Geometry (cs.CG) | 20 | 28 |
| CSAE Working Paper Series | 19 | 28 |
| ADBI Working Papers | 18 | 31 |
| arXiv Neural and Evolutionary Computing (cs.NE) | 18 | 28 |
| Asian Development Bank Economics Working Paper Series | 17 | 27 |
| Brooks World Poverty Institute Working Paper Series | 16 | 35 |
| Instituto de Pesquisa Econômica Aplicada-IPEA Discussion Papers | 14 | 22 |
| Asian Economic Papers | 13 | 26 |
| arXiv History and Philosophy of Physics (physics.hist-ph) | 13 | 18 |
| East Asian Bureau of Economic Research Trade Working Papers | 12 | 25 |
| Escola de Economia de São Paulo, Getulio Vargas Foundation (Brazil) Textos para discussão | 12 | 20 |
| arXiv Graphics (cs.GR) | 9 | 17 |

# APPENDIX 3

**Maximum and minimum h5-index and h5-median values by subject area and subcategory, as they appear in the Google Scholar Metrics publication rankings (2009-2013)**

| AREAS AND SUBCATEGORIES | MAXIMUM H5-INDEX | MINIMUM H5-INDEX | MAXIMUM H5-MEDIAN | MINIMUM H5-MEDIAN |
|---|---|---|---|---|
| **Chemical & Material Sciences** | **193** | **102** | **339** | **133** |
| Analytical Chemistry | 102 | 45 | 132 | 57 |
| Chemical & Material Sciences (general) | 193 | 42 | 339 | 55 |
| Materials Engineering | 174 | 72 | 253 | 95 |
| Nanotechnology | 174 | 32 | 247 | 41 |
| Organic Chemistry | 171 | 37 | 247 | 49 |
| Biochemistry | 164 | 63 | 281 | 79 |
| Electrochemistry | 100 | 10 | 130 | 13 |
| Dispersion Chemistry | 92 | 18 | 112 | 24 |
| Polymers & Plastics | 87 | 31 | 161 | 40 |
| Inorganic Chemistry | 78 | 12 | 114 | 15 |
| Medicinal Chemistry | 78 | 25 | 97 | 30 |
| Chemical Kinetics & Catalysis | 74 | 20 | 97 | 28 |
| Crystallography & Structural Chemistry | 70 | 16 | 85 | 20 |
| Molecular Modeling | 68 | 13 | 97 | 20 |
| Corrosion | 67 | 3 | 89 | 4 |
| Composite Materials | 59 | 11 | 75 | 15 |
| Combustion & Propulsion | 58 | 11 | 97 | 15 |
| Ceramic Engineering | 46 | 7 | 58 | 9 |
| Oil, Petroleum & Natural Gas | 36 | 10 | 44 | 13 |
| **Engineering & Computer Science** | **174** | **93** | **253** | **130** |
| Materials Engineering | 174 | 72 | 253 | 95 |
| Nanotechnology | 174 | 32 | 247 | 41 |
| Biotechnology | 129 | 36 | 197 | 49 |
| Biomedical Technology | 118 | 30 | 143 | 40 |
| Computer Vision & Pattern Recognition | 118 | 25 | 182 | 34 |
| Sustainable Energy | 108 | 29 | 161 | 41 |
| Bioinformatics & Computational Biology | 104 | 19 | 205 | 24 |
| Engineering & Computer Science (general) | 98 | 49 | 162 | 66 |
| Information Theory | 93 | 8 | 152 | 10 |
| Artificial Intelligence | 89 | 36 | 122 | 46 |
| Automation & Control Theory | 86 | 26 | 124 | 35 |
| Computer Networks & Wireless Communication | 86 | 42 | 143 | 58 |
| Signal Processing | 84 | 31 | 127 | 41 |
| Databases & Information Systems | 83 | 32 | 131 | 47 |
| Food Science & Technology | 83 | 35 | 112 | 41 |
| Water Supply & Treatment | 82 | 16 | 105 | 19 |
| Educational Technology | 81 | 18 | 116 | 28 |
| Operations Research | 81 | 27 | 109 | 38 |
| Human Computer Interaction | 78 | 24 | 123 | 28 |
| Power Engineering | 78 | 22 | 104 | 31 |
| Remote Sensing | 75 | 12 | 98 | 15 |
| Computer Hardware Design | 70 | 23 | 94 | 33 |
| Data Mining & Analysis | 69 | 13 | 113 | 19 |
| Computer Graphics | 68 | 9 | 95 | 14 |
| Molecular Modeling | 68 | 13 | 97 | 20 |
| Computer Security & Cryptography | 65 | 27 | 110 | 40 |
| Computational Linguistics | 62 | 15 | 84 | 17 |
| Robotics | 61 | 23 | 88 | 29 |
| Combustion & Propulsion | 58 | 11 | 97 | 15 |
| Evolutionary Computation | 57 | 14 | 92 | 17 |
| Fuzzy Systems | 57 | 12 | 92 | 14 |
| Software Systems | 57 | 27 | 94 | 38 |
| Library & Information Science | 56 | 17 | 80 | 21 |
| Microelectronics & Electronic Packaging | 55 | 19 | 73 | 23 |
| Computing Systems | 54 | 34 | 98 | 49 |
| Environmental & Geological Engineering | 54 | 19 | 72 | 22 |
| Structural Engineering | 54 | 20 | 78 | 25 |
| Metallurgy | 53 | 13 | 66 | 15 |
| Manufacturing & Machinery | 51 | 17 | 72 | 23 |
| Medical Informatics | 51 | 14 | 78 | 17 |
| Theoretical Computer Science | 51 | 24 | 77 | 33 |
| Civil Engineering | 49 | 20 | 62 | 25 |
| Mechanical Engineering | 49 | 22 | 60 | 28 |
| Game Theory and Decision Science | 48 | 16 | 63 | 24 |
| Plasma & Fusion | 48 | 9 | 65 | 11 |
| Transportation | 48 | 24 | 64 | 30 |
| Ceramic Engineering | 46 | 7 | 58 | 9 |
| Multimedia | 46 | 16 | 62 | 22 |
| Quality & Reliability | 44 | 12 | 60 | 14 |
| Wood Science & Technology | 40 | 6 | 52 | 8 |
| Mining & Mineral Resources | 36 | 8 | 47 | 10 |
| Oil, Petroleum & Natural Gas | 36 | 10 | 44 | 13 |
| Textile Engineering | 36 | 4 | 48 | 6 |
| Radar, Positioning & Navigation | 35 | 6 | 52 | 7 |
| Aviation & Aerospace Engineering | 33 | 12 | 46 | 19 |
| Ocean & Marine Engineering | 30 | 11 | 38 | 15 |
| Technology Law | 20 | 7 | 31 | 8 |
| Architecture | 17 | 3 | 25 | 4 |
| **Health & Medical Sciences** | **329** | **129** | **495** | **176** |
| Health & Medical Sciences (general) | 329 | 70 | 495 | 102 |
| Molecular Biology | 223 | 60 | 343 | 78 |
| Oncology | 205 | 66 | 306 | 92 |
| Genetics & Genomics | 188 | 46 | 270 | 59 |
| Cardiology | 178 | 50 | 274 | 62 |
| Hematology | 156 | 34 | 200 | 43 |
| Neurology | 135 | 65 | 220 | 84 |
| Gastroenterology & Hepatology | 132 | 47 | 181 | 62 |
| Immunology | 129 | 48 | 226 | 67 |
| Biomedical Technology | 118 | 30 | 143 | 40 |
| Pediatric Medicine | 117 | 37 | 162 | 44 |
| Pulmonology | 114 | 31 | 150 | 41 |
| Communicable Diseases | 113 | 49 | 148 | 62 |
| Diabetes | 113 | 25 | 163 | 32 |
| Pharmacology & Pharmacy | 110 | 55 | 186 | 68 |
| Rheumatology | 107 | 26 | 148 | 32 |
| Endocrinology | 106 | 46 | 139 | 60 |
| Nutrition Science | 103 | 34 | 136 | 46 |
| Psychiatry | 103 | 50 | 143 | 67 |
| Vascular Medicine | 103 | 33 | 136 | 44 |
| Virology | 98 | 32 | 125 | 39 |
| Gynecology & Obstetrics | 97 | 33 | 175 | 40 |
| Urology & Nephrology | 96 | 36 | 144 | 46 |
| Critical Care | 95 | 13 | 140 | 17 |
| Psychology | 92 | 52 | 142 | 67 |
| Radiology & Medical Imaging | 92 | 39 | 122 | 53 |
| Surgery | 90 | 36 | 124 | 48 |
| Toxicology | 89 | 39 | 118 | 48 |
| Physiology | 83 | 34 | 179 | 46 |
| AIDS & HIV | 82 | 15 | 122 | 19 |
| Anesthesiology | 82 | 22 | 115 | 28 |
| Nuclear Medicine, Radiotherapy & Molecular Imaging | 80 | 26 | 107 | 32 |
| Medicinal Chemistry | 78 | 25 | 97 | 30 |
| Ophthalmology & Optometry | 78 | 34 | 110 | 39 |
| Orthopedic Medicine & Surgery | 75 | 38 | 100 | 49 |
| Physical Education & Sports Medicine | 75 | 36 | 95 | 46 |
| Reproductive Health | 75 | 29 | 103 | 36 |
| Social Psychology | 75 | 21 | 114 | 31 |
| Clinical Laboratory Science | 74 | 10 | 109 | 11 |
| Epidemiology | 74 | 29 | 97 | 38 |
| Transplantation | 74 | 13 | 100 | 18 |
| Pathology | 73 | 29 | 119 | 36 |
| Pain & Pain Management | 72 | 13 | 106 | 19 |
| Gerontology & Geriatric Medicine | 71 | 36 | 99 | 48 |
| Dermatology | 70 | 28 | 95 | 35 |
| Obesity | 69 | 6 | 95 | 7 |
| Public Health | 69 | 31 | 98 | 39 |
| Child & Adolescent Psychology | 64 | 30 | 95 | 40 |
| Dentistry | 63 | 31 | 81 | 40 |
| Heart & Thoracic Surgery | 62 | 10 | 79 | 13 |
| Addiction | 61 | 21 | 84 | 26 |
| Developmental Disabilities | 59 | 16 | 78 | 25 |
| Natural Medicines & Medicinal Plants | 59 | 20 | 87 | 23 |
| Plastic & Reconstructive Surgery | 59 | 12 | 72 | 13 |
| Tropical Medicine & Parasitology | 59 | 25 | 74 | 27 |
| Emergency Medicine | 58 | 18 | 85 | 21 |
| Primary Health Care | 58 | 11 | 84 | 16 |
| Neurosurgery | 57 | 13 | 74 | 16 |
| Veterinary Medicine | 54 | 24 | 73 | 29 |
| Medical Informatics | 51 | 14 | 78 | 17 |
| Nursing | 49 | 24 | 67 | 29 |
| Rehabilitation Therapy | 48 | 20 | 60 | 32 |
| Otolaryngology | 47 | 23 | 66 | 27 |
| Hospice & Palliative Care | 46 | 7 | 62 | 9 |
| Oral & Maxillofacial Surgery | 44 | 13 | 62 | 16 |
| Pregnancy & Childbirth | 39 | 21 | 58 | 31 |
| Alternative & Traditional Medicine | 37 | 16 | 54 | 21 |
| Audiology, Speech & Language Pathology | 35 | 11 | 51 | 17 |
| Bioethics | 29 | 8 | 40 | 11 |
| **Life Sciences & Earth Sciences** | **355** | **108** | **495** | **146** |
| Life Sciences & Earth Sciences (general) | 355 | 57 | 495 | 78 |
| Molecular Biology | 223 | 60 | 343 | 78 |
| Cell Biology | 223 | 72 | 343 | 95 |
| Biochemistry | 164 | 63 | 281 | 79 |
| Biotechnology | 129 | 36 | 197 | 49 |
| Developmental Biology & Embryology | 121 | 29 | 192 | 35 |
| Environmental Sciences | 115 | 44 | 151 | 55 |
| Sustainable Energy | 108 | 29 | 161 | 41 |
| Microbiology | 105 | 47 | 161 | 60 |
| Bioinformatics & Computational Biology | 104 | 19 | 205 | 24 |
| Virology | 98 | 32 | 125 | 39 |
| Botany | 93 | 42 | 132 | 51 |
| Atmospheric Sciences | 89 | 34 | 120 | 46 |
| Evolutionary Biology | 86 | 18 | 142 | 21 |
| Food Science & Technology | 83 | 35 | 112 | 41 |
| Ecology | 83 | 37 | 124 | 53 |
| Sustainable Development | 75 | 25 | 105 | 34 |
| Proteomics, Peptides & Aminoacids | 74 | 20 | 101 | 27 |
| Biophysics | 65 | 25 | 81 | 30 |
| Hydrology & Water Resources | 62 | 15 | 83 | 20 |
| Soil Sciences | 61 | 21 | 82 | 27 |
| Biodiversity & Conservation Biology | 60 | 28 | 89 | 35 |
| Geology | 59 | 30 | 74 | 40 |
| Geochemistry & Mineralogy | 58 | 16 | 71 | 23 |
| Animal Husbandry | 56 | 19 | 70 | 23 |
| Agronomy & Crop Science | 55 | 30 | 75 | 40 |
| Environmental & Geological Engineering | 54 | 19 | 72 | 22 |
| Forests & Forestry | 53 | 15 | 72 | 17 |
| Oceanography | 49 | 30 | 84 | 34 |
| Marine Sciences & Fisheries | 49 | 33 | 71 | 42 |
| Zoology | 48 | 18 | 64 | 23 |
| Insects & Arthropods | 48 | 20 | 81 | 27 |
| Animal Behavior & Ethology | 47 | 11 | 65 | 13 |
| Plant Pathology | 46 | 13 | 75 | 16 |
| Wood Science & Technology | 40 | 6 | 52 | 8 |
| Paleontology | 39 | 12 | 50 | 16 |
| Pest Control & Pesticides | 37 | 10 | 46 | 12 |
| Mycology | 37 | 11 | 49 | 16 |
| Birds | 25 | 7 | 36 | 10 |
| **Physics & Mathematics** | **191** | **107** | **263** | **140** |
| Physics & Mathematics (general) | 191 | 43 | 263 | 54 |
| Astronomy & Astrophysics | 162 | 33 | 222 | 44 |
| High Energy & Nuclear Physics | 145 | 39 | 194 | 47 |
| Condensed Matter Physics & Semiconductors | 138 | 38 | 227 | 48 |
| Quantum Mechanics | 128 | 7 | 204 | 9 |
| Optics & Photonics | 122 | 35 | 186 | 53 |
| Geophysics | 91 | 27 | 127 | 36 |
| Spectroscopy & Molecular Physics | 73 | 26 | 101 | 33 |
| Biophysics | 65 | 25 | 81 | 30 |
| Mathematical Analysis | 64 | 22 | 95 | 30 |
| Pure & Applied Mathematics | 64 | 16 | 95 | 19 |
| Mathematical Optimization | 61 | 18 | 99 | 24 |
| Nonlinear Science | 61 | 21 | 78 | 26 |
| Probability & Statistics with Applications | 61 | 34 | 113 | 39 |
| Computational Mathematics | 60 | 25 | 78 | 34 |
| Electromagnetism | 58 | 20 | 75 | 25 |
| Thermal Sciences | 58 | 17 | 76 | 20 |
| Mathematical Physics | 52 | 15 | 78 | 20 |
| Fluid Mechanics | 51 | 14 | 77 | 16 |
| Acoustics & Sound | 49 | 16 | 75 | 23 |
| Plasma & Fusion | 48 | 9 | 65 | 11 |
| Algebra | 43 | 20 | 62 | 26 |
| Geometry | 43 | 20 | 62 | 27 |
| Discrete Mathematics | 37 | 18 | 48 | 22 |
| **Business, Economics & Management** | **168** | **63** | **241** | **92** |
| Economics | 168 | 50 | 241 | 80 |
| Finance | 116 | 31 | 179 | 41 |
| Economic Policy | 105 | 32 | 148 | 53 |
| Human Resources & Organizations | 81 | 37 | 118 | 54 |
| Entrepreneurship & Innovation | 73 | 19 | 101 | 27 |
| Business, Economics & Management (general) | 72 | 33 | 122 | 52 |
| Strategic Management | 72 | 42 | 113 | 61 |
| Marketing | 65 | 26 | 111 | 36 |
| Tourism & Hospitality | 63 | 13 | 91 | 18 |
| Development Economics | 62 | 16 | 111 | 20 |
| International Business | 60 | 15 | 84 | 24 |
| Accounting & Taxation | 57 | 20 | 87 | 28 |
| Game Theory and Decision Science | 48 | 16 | 63 | 24 |
| Emergency Management | 39 | 4 | 56 | 7 |
| Educational Administration | 32 | 11 | 48 | 14 |
| Economic History | 22 | 3 | 33 | 3 |
| **Social Sciences** | **81** | **45** | **116** | **59** |
| Cognitive Science | 92 | 32 | 138 | 42 |
| Environmental & Occupational Medicine | 89 | 28 | 118 | 34 |
| Environmental Law & Policy | 87 | 16 | 115 | 20 |
| Educational Technology | 81 | 18 | 116 | 28 |
| Human Resources & Organizations | 81 | 37 | 118 | 54 |
| Health Policy & Medical Law | 77 | 9 | 111 | 11 |
| Sustainable Development | 75 | 25 | 105 | 34 |
| Ethics | 70 | 10 | 98 | 11 |
| Public Health | 69 | 31 | 98 | 39 |
| Social Sciences (general) | 66 | 31 | 96 | 38 |
| Development Economics | 62 | 16 | 111 | 20 |
| Library & Information Science | 56 | 17 | 80 | 21 |
| Political Science | 54 | 25 | 85 | 33 |
| Education | 52 | 32 | 86 | 46 |
| Educational Psychology & Counseling | 52 | 17 | 74 | 22 |
| Science & Engineering Education | 52 | 20 | 70 | 28 |
| Diplomacy & International Relations | 48 | 16 | 75 | 22 |
| Geography & Cartography | 47 | 18 | 70 | 24 |
| Sociology | 47 | 30 | 82 | 38 |
| Teaching & Teacher Education | 47 | 18 | 61 | 22 |
| Sex & Sexuality | 46 | 11 | 60 | 13 |
| Urban Studies & Planning | 46 | 24 | 79 | 34 |
| Family Studies | 44 | 13 | 64 | 15 |
| Public Policy & Administration | 42 | 15 | 57 | 22 |
| Academic & Psychological Testing | 40 | 10 | 54 | 13 |
| Anthropology | 40 | 12 | 51 | 15 |
| Forensic Science | 40 | 10 | 58 | 15 |
| Higher Education | 40 | 15 | 54 | 21 |
| Law | 39 | 27 | 71 | 37 |
| Paleontology | 39 | 12 | 50 | 16 |
| Social Work | 39 | 21 | 60 | 27 |
| Archaeology | 38 | 10 | 45 | 14 |
| Criminology, Criminal Law & Policing | 36 | 19 | 49 | 25 |
| Special Education | 33 | 16 | 48 | 23 |
| Educational Administration | 32 | 11 | 48 | 14 |
| Epistemology & Scientific History | 32 | 9 | 43 | 12 |
| Human Migration | 32 | 9 | 55 | 13 |
| Early Childhood Education | 30 | 12 | 59 | 19 |
| Bioethics | 29 | 8 | 40 | 11 |
| Chinese Studies & History | 29 | 6 | 44 | 9 |
| Military Studies | 27 | 9 | 45 | 11 |
| Feminism & Women's Studies | 25 | 6 | 36 | 9 |
| African Studies & History | 24 | 8 | 38 | 12 |
| Asian Studies & History | 24 | 12 | 36 | 16 |
| International Law | 23 | 14 | 40 | 19 |
| Economic History | 22 | 3 | 33 | 3 |
| History | 22 | 8 | 33 | 12 |
| European Law | 21 | 5 | 31 | 5 |
| Technology Law | 20 | 7 | 31 | 8 |
| Architecture | 17 | 3 | 25 | 4 |
| Canadian Studies & History | 17 | 3 | 26 | 5 |
| Middle Eastern & Islamic Studies | 15 | 5 | 24 | 6 |
| **Humanities, Literature & Arts** | **38** | **25** | **72** | **30** |
| Sex & Sexuality | 46 | 11 | 60 | 13 |
| Communication | 45 | 20 | 72 | 31 |
| Language & Linguistics | 36 | 16 | 51 | 23 |
| Gender Studies | 35 | 10 | 41 | 12 |
| Foreign Language Learning | 34 | 13 | 50 | 18 |
| Epistemology & Scientific History | 32 | 9 | 43 | 12 |
| Ethnic & Cultural Studies | 32 | 10 | 55 | 14 |
| Philosophy | 32 | 15 | 44 | 20 |
| Humanities, Literature & Arts (general) | 31 | 14 | 44 | 18 |
| Music & Musicology | 31 | 9 | 43 | 12 |
| Chinese Studies & History | 29 | 6 | 44 | 9 |
| Religion | 26 | 8 | 30 | 11 |
| Feminism & Women's Studies | 25 | 6 | 36 | 9 |
| African Studies & History | 24 | 8 | 38 | 12 |
| Asian Studies & History | 24 | 12 | 36 | 16 |
| English Language & Literature | 23 | 5 | 35 | 6 |
| History | 22 | 8 | 33 | 12 |
| Canadian Studies & History | 17 | 3 | 26 | 5 |
| Latin American Studies | 17 | 4 | 27 | 5 |
| Middle Eastern & Islamic Studies | 15 | 5 | 24 | 6 |
| Visual Arts | 13 | 4 | 18 | 5 |
| Drama & Theater Arts | 12 | 4 | 17 | 4 |
| Literature & Writing | 11 | 6 | 24 | 7 |
| Film | 10 | 4 | 19 | 5 |
| American Literature & Studies | 8 | 3 | 15 | 4 |
| French Studies | 8 | 1 | 12 | 2 |

# APENDIX 4
## Correlation between the positions of the publications by languages in the last two editions of GSM (2008-2012 and 2009-2013)

| SUBCATEGORIES | CORRELATION |
|---|---|
| Dutch | 0.91 |
| German | 0.88 |
| Portuguese | 0.77 |
| Spanish | 0.77 |
| Chinese | 0.77 |
| Italian | 0.68 |
| French | 0.64 |
| Japanese | 0.62 |

## APPENDIX 5
### Correlation between the positions of the publications by subcategories in the last two editions of GSM (2008-2012 and 2009-2013)

| SUBCATEGORIES | CORRELATION | ADDITIONS / DELETIONS % |
|---|---|---|
| **Business, Economics & Management** | | |
| Emergency Management | 0,98 | 20 |
| Economic History | 0,97 | 25 |
| Marketing | 0,96 | 10 |
| Human Resources & Organizations | 0,95 | 10 |
| Entrepreneurship & Innovation | 0,95 | 5 |
| International Business | 0,95 | 5 |
| Economics | 0,94 | 5 |
| Finance | 0,94 | 0 |
| Development Economics | 0,94 | 5 |
| Game Theory and Decision Science | 0,93 | 15 |
| Educational Administration | 0,93 | 15 |
| Accounting & Taxation | 0,92 | 10 |
| Tourism & Hospitality | 0,92 | 10 |
| Strategic Management | 0,89 | 5 |
| Economic Policy | 0,86 | 10 |
| Business, Economics & Management (general) | 0,80 | 20 |
| **Chemical & Material Sciences** | | |
| Corrosion | 0,99 | 15 |
| Electrochemistry | 0,98 | 10 |
| Inorganic Chemistry | 0,98 | 25 |
| Composite Materials | 0,98 | 5 |
| Dispersion Chemistry | 0,97 | 5 |
| Materials Engineering | 0,96 | 5 |
| Analytical Chemistry | 0,96 | 5 |
| Nanotechnology | 0,96 | 5 |
| Chemical Kinetics & Catalysis | 0,95 | 10 |
| Medicinal Chemistry | 0,95 | 5 |
| Polymers & Plastics | 0,95 | 15 |
| Ceramic Engineering | 0,95 | 15 |
| Oil, Petroleum & Natural Gas | 0,93 | 5 |
| Crystallography & Structural Chemistry | 0,92 | 10 |
| Combustion & Propulsion | 0,92 | 0 |
| Biochemistry | 0,91 | 25 |
| Organic Chemistry | 0,88 | 35 |
| Chemical & Material Sciences (general) | 0,87 | 50 |
| Molecular Modeling | 0,85 | 5 |

| Engineering & Computer Science | | |
|---|---|---|
| Metallurgy | 0,99 | 10 |
| Mining & Mineral Resources | 0,98 | 30 |
| Plasma & Fusion | 0,98 | 0 |
| Data Mining & Analysis | 0,97 | 20 |
| Fuzzy Systems | 0,97 | 30 |
| Remote Sensing | 0,97 | 10 |
| Technology Law | 0,97 | 25 |
| Sustainable Energy | 0,97 | 15 |
| Structural Engineering | 0,96 | 5 |
| Ocean & Marine Engineering | 0,96 | 10 |
| Materials Engineering | 0,96 | 5 |
| Water Supply & Treatment | 0,96 | 10 |
| Computer Graphics | 0,96 | 10 |
| Robotics | 0,96 | 10 |
| Nanotechnology | 0,96 | 5 |
| Wood Science & Technology | 0,95 | 5 |
| Bioinformatics & Computational Biology | 0,95 | 20 |
| Power Engineering | 0,95 | 20 |
| Medical Informatics | 0,95 | 20 |
| Information Theory | 0,95 | 10 |
| Ceramic Engineering | 0,95 | 15 |
| Computer Hardware Design | 0,95 | 5 |
| Computer Vision & Pattern Recognition | 0,95 | 15 |
| Software Systems | 0,94 | 20 |
| Library & Information Science | 0,94 | 15 |
| Engineering & Computer Science (general) | 0,94 | 15 |
| Oil, Petroleum & Natural Gas | 0,93 | 5 |
| Game Theory and Decision Science | 0,93 | 15 |
| Textile Engineering | 0,93 | 20 |
| Biotechnology | 0,93 | 15 |
| Combustion & Propulsion | 0,92 | 0 |
| Radar, Positioning & Navigation | 0,92 | 10 |
| Transportation | 0,92 | 15 |
| Computer Security & Cryptography | 0,91 | 25 |
| Computing Systems | 0,91 | 35 |
| Automation & Control Theory | 0,91 | 0 |
| Manufacturing & Machinery | 0,91 | 5 |
| Computer Networks & Wireless Communication | 0,91 | 10 |
| Signal Processing | 0,90 | 0 |
| Operations Research | 0,90 | 0 |
| Civil Engineering | 0,90 | 5 |
| Evolutionary Computation | 0,89 | 20 |
| Biomedical Technology | 0,89 | 35 |
| Theoretical Computer Science | 0,89 | 15 |
| Databases & Information Systems | 0,89 | 20 |
| Mechanical Engineering | 0,88 | 10 |
| Human Computer Interaction | 0,88 | 30 |
| Artificial Intelligence | 0,87 | 15 |
| Quality & Reliability | 0,86 | 15 |
| Molecular Modeling | 0,85 | 5 |
| Educational Technology | 0,85 | 20 |
| Aviation & Aerospace Engineering | 0,84 | 10 |
| Microelectronics & Electronic Packaging | 0,83 | 10 |
| Food Science & Technology | 0,81 | 10 |
| Environmental & Geological Engineering | 0,81 | 10 |
| Multimedia | 0,80 | 20 |
| Computational Linguistics | 0,80 | 15 |
| Architecture | 0,77 | 25 |
| **Health & Medical Sciences** | | |
| Pain & Pain Management | 0,99 | 10 |
| Vascular Medicine | 0,99 | 5 |
| Obesity | 0,98 | 5 |
| Anesthesiology | 0,98 | 5 |
| Oncology | 0,98 | 10 |
| Pathology | 0,98 | 20 |
| Surgery | 0,98 | 10 |
| Transplantation | 0,98 | 0 |
| AIDS & HIV | 0,98 | 0 |
| Molecular Biology | 0,98 | 15 |
| Critical Care | 0,98 | 10 |
| Pulmonology | 0,98 | 10 |
| Hospice & Palliative Care | 0,97 | 5 |
| Addiction | 0,97 | 10 |
| Clinical Laboratory Science | 0,97 | 10 |
| Primary Health Care | 0,97 | 5 |
| Bioethics | 0,97 | 5 |
| Physiology | 0,97 | 10 |
| Health & Medical Sciences (general) | 0,97 | 25 |
| Communicable Diseases | 0,96 | 5 |
| Diabetes | 0,96 | 5 |
| Epidemiology | 0,96 | 5 |
| Genetics & Genomics | 0,96 | 15 |
| Audiology, Speech & Language Pathology | 0,96 | 10 |
| Immunology | 0,96 | 5 |
| Psychiatry | 0,96 | 10 |
| Cardiology | 0,96 | 20 |
| Nutrition Science | 0,96 | 15 |
| Medicinal Chemistry | 0,95 | 5 |
| Gastroenterology & Hepatology | 0,95 | 5 |
| Neurosurgery | 0,95 | 5 |
| Toxicology | 0,95 | 20 |
| Rheumatology | 0,95 | 5 |
| Emergency Medicine | 0,95 | 15 |
| Pediatric Medicine | 0,95 | 10 |
| Medical Informatics | 0,95 | 20 |
| Oral & Maxillofacial Surgery | 0,95 | 10 |
| Urology & Nephrology | 0,95 | 15 |
| Neurology | 0,94 | 0 |
| Ophthalmology & Optometry | 0,94 | 5 |
| Pregnancy & Childbirth | 0,94 | 10 |
| Rehabilitation Therapy | 0,94 | 15 |
| Plastic & Reconstructive Surgery | 0,94 | 5 |
| Natural Medicines & Medicinal Plants | 0,94 | 20 |
| Social Psychology | 0,93 | 15 |
| Hematology | 0,93 | 10 |
| Physical Education & Sports Medicine | 0,93 | 5 |
| Dermatology | 0,93 | 15 |
| Reproductive Health | 0,93 | 5 |
| Veterinary Medicine | 0,91 | 35 |
| Heart & Thoracic Surgery | 0,91 | 0 |
| Tropical Medicine & Parasitology | 0,91 | 10 |
| Gynecology & Obstetrics | 0,91 | 5 |
| Psychology | 0,91 | 15 |
| Public Health | 0,91 | 15 |
| Nuclear Medicine, Radiotherapy & Molecular Imaging | 0,90 | 15 |
| Orthopedic Medicine & Surgery | 0,90 | 15 |
| Biomedical Technology | 0,89 | 35 |
| Developmental Disabilities | 0,89 | 5 |
| Alternative & Traditional Medicine | 0,89 | 10 |
| Virology | 0,88 | 25 |
| Endocrinology | 0,88 | 20 |
| Dentistry | 0,87 | 10 |
| Child & Adolescent Psychology | 0,87 | 10 |
| Radiology & Medical Imaging | 0,86 | 0 |
| Otolaryngology | 0,86 | 5 |
| Pharmacology & Pharmacy | 0,86 | 25 |
| Nursing | 0,86 | 20 |
| Gerontology & Geriatric Medicine | 0,82 | 5 |
| **Humanities, Literature & Arts** | | |
| Gender Studies | 0,97 | 10 |
| Sex & Sexuality | 0,96 | 20 |
| Foreign Language Learning | 0,96 | 5 |
| Chinese Studies & History | 0,95 | 10 |
| Epistemology & Scientific History | 0,95 | 25 |
| Music & Musicology | 0,94 | 5 |
| Middle Eastern & Islamic Studies | 0,93 | 10 |
| Canadian Studies & History | 0,92 | 45 |
| Drama & Theater Arts | 0,92 | 25 |
| History | 0,91 | 25 |
| Latin American Studies | 0,91 | 15 |
| Language & Linguistics | 0,90 | 10 |
| Visual Arts | 0,89 | 35 |
| Ethnic & Cultural Studies | 0,88 | 5 |
| English Language & Literature | 0,84 | 10 |
| African Studies & History | 0,83 | 10 |
| French Studies | 0,81 | 20 |
| Feminism & Women's Studies | 0,81 | 10 |
| Religion | 0,78 | 15 |
| Philosophy | 0,76 | 10 |
| Asian Studies & History | 0,74 | 10 |
| American Literature & Studies | 0,73 | 25 |
| Communication | 0,69 | 10 |
| Humanities, Literature & Arts (general) | 0,66 | 20 |
| Film | 0,54 | 25 |
| Literature & Writing | 0,32 | 20 |
| **Life Sciences & Earth Sciences** | | |
| Biodiversity & Conservation Biology | 0,98 | 20 |
| Molecular Biology | 0,98 | 15 |
| Environmental Sciences | 0,98 | 10 |
| Birds | 0,97 | 10 |
| Hydrology & Water Resources | 0,97 | 0 |
| Pest Control & Pesticides | 0,97 | 10 |
| Sustainable Energy | 0,97 | 15 |
| Mycology | 0,96 | 15 |
| Animal Behavior & Ethology | 0,96 | 20 |
| Life Sciences & Earth Sciences (general) | 0,96 | 35 |
| Plant Pathology | 0,96 | 15 |
| Wood Science & Technology | 0,95 | 5 |
| Evolutionary Biology | 0,95 | 15 |
| Paleontology | 0,95 | 5 |
| Bioinformatics & Computational Biology | 0,95 | 20 |
| Developmental Biology & Embryology | 0,95 | 5 |
| Proteomics, Peptides & Aminoacids | 0,95 | 5 |
| Biophysics | 0,94 | 15 |
| Cell Biology | 0,94 | 10 |
| Geology | 0,94 | 15 |
| Soil Sciences | 0,94 | 5 |
| Botany | 0,94 | 10 |
| Sustainable Development | 0,94 | 10 |
| Forests & Forestry | 0,94 | 5 |
| Biotechnology | 0,93 | 15 |
| Animal Husbandry | 0,93 | 10 |
| Geochemistry & Mineralogy | 0,93 | 15 |
| Atmospheric Sciences | 0,93 | 15 |
| Virology | 0,93 | 10 |
| Ecology | 0,92 | 10 |
| Marine Sciences & Fisheries | 0,91 | 10 |
| Biochemistry | 0,91 | 25 |
| Zoology | 0,90 | 15 |
| Insects & Arthropods | 0,89 | 10 |
| Microbiology | 0,88 | 5 |
| Food Science & Technology | 0,81 | 10 |
| Environmental & Geological Engineering | 0,81 | 10 |
| Oceanography | 0,79 | 10 |
| Agronomy & Crop Science | 0,73 | 20 |
| **Physics & Mathematics** | | |
| Mathematical Physics | 0,99 | 5 |
| Fluid Mechanics | 0,98 | 20 |
| Plasma & Fusion | 0,98 | 0 |
| Mathematical Optimization | 0,98 | 10 |
| Quantum Mechanics | 0,98 | 0 |
| Astronomy & Astrophysics | 0,97 | 5 |
| Probability & Statistics with Applications | 0,97 | 15 |
| Mathematical Analysis | 0,97 | 5 |
| Computational Mathematics | 0,97 | 0 |
| Algebra | 0,96 | 10 |
| High Energy & Nuclear Physics | 0,95 | 10 |
| Acoustics & Sound | 0,95 | 15 |
| Thermal Sciences | 0,95 | 5 |
| Discrete Mathematics | 0,95 | 5 |
| Nonlinear Science | 0,95 | 15 |
| Biophysics | 0,94 | 15 |
| Condensed Matter Physics & Semiconductors | 0,94 | 10 |
| Optics & Photonics | 0,94 | 10 |
| Physics & Mathematics (general) | 0,93 | 15 |
| Spectroscopy & Molecular Physics | 0,92 | 15 |
| Geometry | 0,91 | 10 |
| Pure & Applied Mathematics | 0,90 | 15 |
| Geophysics | 0,83 | 25 |
| Electromagnetism | 0,70 | 10 |
| **Social Sciences** | | |
| Health Policy & Medical Law | 0,98 | 20 |
| Economic History | 0,97 | 25 |
| Geography & Cartography | 0,97 | 15 |
| Academic & Psychological Testing | 0,97 | 15 |
| Technology Law | 0,97 | 25 |
| Diplomacy & International Relations | 0,97 | 25 |
| Sex & Sexuality | 0,96 | 20 |
| Education | 0,96 | 10 |
| Forensic Science | 0,96 | 10 |
| European Law | 0,96 | 30 |
| Ethics | 0,96 | 15 |
| Human Resources & Organizations | 0,95 | 10 |
| Chinese Studies & History | 0,95 | 10 |
| Epistemology & Scientific History | 0,95 | 25 |
| Paleontology | 0,95 | 5 |
| International Law | 0,95 | 15 |
| Political Science | 0,95 | 5 |
| Early Childhood Education | 0,95 | 10 |
| Urban Studies & Planning | 0,94 | 10 |
| Educational Psychology & Counseling | 0,94 | 0 |
| Library & Information Science | 0,94 | 15 |
| Environmental Law & Policy | 0,94 | 10 |
| Sustainable Development | 0,94 | 10 |
| Development Economics | 0,94 | 5 |
| Anthropology | 0,93 | 25 |
| Educational Administration | 0,93 | 15 |
| Middle Eastern & Islamic Studies | 0,93 | 10 |
| Human Migration | 0,93 | 0 |
| Family Studies | 0,93 | 15 |
| Canadian Studies & History | 0,92 | 45 |
| Archaeology | 0,92 | 15 |
| Military Studies | 0,91 | 5 |
| History | 0,91 | 25 |
| Cognitive Science | 0,91 | 10 |
| Higher Education | 0,91 | 10 |
| Teaching & Teacher Education | 0,91 | 15 |
| Public Health | 0,91 | 15 |
| Science & Engineering Education | 0,90 | 5 |
| Environmental & Occupational Medicine | 0,90 | 15 |
| Criminology, Criminal Law & Policing | 0,88 | 15 |
| Public Policy & Administration | 0,87 | 10 |
| Special Education | 0,87 | 25 |
| Educational Technology | 0,85 | 20 |
| Law | 0,84 | 5 |
| African Studies & History | 0,83 | 10 |
| Feminism & Women's Studies | 0,81 | 10 |
| Social Work | 0,78 | 20 |
| Architecture | 0,77 | 25 |
| Asian Studies & History | 0,74 | 10 |
| Social Sciences (general) | 0,56 | 35 |